\documentclass[prc,aps,nofootinbib,superscriptaddress,showkeys,showpacs,twocolumn,10pt,floatfix]{revtex4-1}
\usepackage{epsfig}
\usepackage{graphicx}
\usepackage{amsmath,amssymb,amsfonts}
\usepackage[small]{subfigure}
\usepackage[dvipsnames]{color}
\begin{document}
\title{Lee-Yang--inspired energy--density functional including \\ contributions from $p$--wave scattering}
\author{J{\'e}r{\'e}my Bonnard}
\email{jeremy.bonnard@york.ac.uk}
\affiliation{Department of Physics, University of York, Heslington, York YO10 5DD, United Kingdom}
\affiliation{Universit\'e Paris-Saclay, CNRS/IN2P3, IJCLab, 91405 Orsay, France}
\author{Marcella Grasso}
\email{grasso@ipno.in2p3.fr}
\affiliation{Universit\'e Paris-Saclay, CNRS/IN2P3, IJCLab, 91405 Orsay, France}
\author{Denis Lacroix}
\email{lacroix@ipno.in2p3.fr}
\affiliation{Universit\'e Paris-Saclay, CNRS/IN2P3, IJCLab, 91405 Orsay, France}
\begin{abstract}
The ELYO functional proposed in [M. Grasso, D. Lacroix, and C. J. Yang, Phys. Rev. C \textbf{95}, 054327 (2017)] belongs to the family of energy-density functionals (EDFs) inspired by effective--field theories and constrained by \textit{ab--initio} pseudo--data. We present here an extension of this EDF which also accounts for the first $p$--wave term appearing in the low--density expansion from which it derives. It is shown that this enrichment of the ansatz on which the functional is based leads to a significant improvement of the description of neutronic systems, especially in regimes besides the pseudo--data set employed to adjust the parameters. As an illustrative application, the mass--radius relation of neutron stars is considered. In contrast to its initial version, the new functional predicts values which are qualitatively consistent with recent observations.
\end{abstract}
\date{\today}
\maketitle
%
\section{Introduction}
%
EDF theories represent a rich and versatile theoretical framework in nuclear physics \cite{bender}. The description they provide for nuclear structure and reactions is globally very satisfactory and covers in practice the whole nuclear chart. However, nuclear EDFs are built on an empirical basis. Several attempts to link them more tightly with microscopic ingredients and theories were carried out in different ways. The past years have seen the emergence of new strategies for developing EDFs (see, for example, Refs. \cite{EFTEDF0,EFTEDF} and Ref. \cite{ppnp2019} for a recent review). The underlying idea is to implement techniques or to adapt results \cite{O2Sk1,O2Sk2,NLOEDF,YGLO,KIDS1,ELYO} from chiral effective-field theories (EFTs) \cite{Bed02,EFTrev2,EFT0,EFT00} and {\it{ab--initio}} models, with the primary aim of rendering the designed EDFs less phenomenological than the traditional ones generated from Skyrme or Gogny effective interactions \cite{sk0,sk1,sk2,gogny,gogny2}.

Our group recently proposed several procedures to reduce the empirical nature of nuclear functionals. The so--called YGLO (Yang-Grasso-Lacroix-Orsay) functional \cite{YGLO} contains a resummed formula, in the same spirit as in EFTs \cite{schafer,kaiser,steele}, to account for the large value of the neutron-neutron scattering length, together with Skyrme--type velocity-- and density--dependent terms, which guarantee the saturation properties of symmetric matter and a correct behavior of neutron matter at all density scales. Some parameters of the resummed term were linked to the neutron-neutron scattering length so to reproduce the Lee-Yang expansion that is valid for very dilute Fermi gases \cite{LY,Bis73}. Only seven parameters remained unconstrained by this requirement and were adjusted on microscopic pseudo--data available for symmetric and neutron matter.

Later, an alternative way was proposed to deal with the large value of the neutron-neutron scattering length and to reproduce at the same time the equation of state (EOS) of neutron matter both in the very dilute regime (Lee-Yang expansion) and close to the saturation density of symmetric matter (density scales of interest for finite nuclei), without resorting to a resummation. A new functional was introduced, containing a density--dependent neutron-neutron scattering length \cite{ELYO}. This functional was later called ELYO (extended Lee-Yang, Orsay) in Ref. \cite{Drops}, where it was generalized for treating finite--size systems. To construct such a functional, a Lee-Yang--inspired EOS was considered. Thus, the functional can be regarded as EFT--inspired in the sense that it correctly describes, by construction, dilute Fermi gases (as is the case in EFT). In addition, such a functional can be regarded as \textit{ab--initio}--inspired in the sense that it was benchmarked on {\it{ab--initio}} pseudodata for reproducing the energies of neutron drops. The validity of the ELYO functional at all neutron--matter densities was guaranteed through the use of a scattering length tuned as a function of the density by imposing a low--density constraint $|a_sk_F|<1$, where $a_s$ is the scattering length and $k_F$ the Fermi momentum.  Satisfactory EOSs were obtained up to the nuclear saturation density for both pure neutron matter (PNM) and symmetric nuclear matter (SNM). In the case of PNM only one parameter was fitted, the effective range $r_s$ associated with the $s$--wave scattering length $a_s$, in order to have a Lee-Yang--type EOS valid at all densities. To describe also SNM, a mapping was carried out with an $s$--wave Skyrme--like EOS and four parameters were adjusted (the Skyrme parameters $t_0$, $t_1$, $t_3$, and $\alpha$, where $\alpha$ is the power of the density--dependent term). 

A recent application to neutron drops \cite{Drops}, finite-size systems composed solely of neutrons living in a harmonic trapping potential,  has revealed severe limitations of the ELYO EDF that may be intuitively understood by analyzing the associated PNM EOS.
As an attempt to overcome the observed drawbacks, we propose here to extend the functional \textit{via} the inclusion of the $p$--wave contribution to the energy, neglected in the initial version.

The present work is organized in the following way. The ELYO ansatz is first enriched for PNM in Sec. \ref{sec_PNM}. Next, neutron drops are addressed in Sec. \ref{sec_drops} where additional parameters defining the effective mass are introduced. In Sec. \ref{sec_SNM}, the remaining free parameters are determined by using the EOS of SNM as constraint, which entirely specifies the proposed functional. As an application, masses and radii of neutron stars are then evaluated in Sec. \ref{sec_NS} by solving the Tolman-Oppenheimer-Volkov equations. Finally, conclusions are given in Sec. \ref{concl}.
%
%
%
\section{Pure neutron matter: Lee-Yang expansion} \label{sec_PNM}
%
Let $a_p$ be the neutron-neutron $p$--wave scattering length. When the density $\rho$ is such that the Fermi momentum $k_F=(3\pi^2\rho)^{1/3}$ satisfy simultaneously the following conditions
\begin{equation} \label{validity}
|a_sk_F| < 1,\quad |r_sk_F| < 1 ,\quad |a_pk_F| < 1,    
\end{equation} 
a low--density expansion may be performed \cite{LY,Bis73}, which was first derived by Lee and Yang in the 1950s and which naturally arises in EFT as shown in Ref. \cite{EFT1}. The first terms of this expansion are 
\begin{align}\label{LY}
\nonumber & \dfrac{E}{N} = \dfrac{\hbar^2 k_F^2}{2m}\biggl\lbrace \dfrac{3}{5} + \dfrac{2}{3\pi}(a_sk_F) + \dfrac{4}{35\pi^2} (11-2\ln2)(a_sk_F)^2 \\ 
& \! +\! \dfrac{1}{10\pi} (r_sk_F)(a_sk_F)^2 \!+ \! 0.019(a_sk_F)^3 +\! \dfrac{3}{5\pi} (a_pk_F)^3\!\biggr\rbrace\!,
\end{align}
where $N$ and $m$ denote, respectively, the neutron number and the neutron mass.

The values of the physical constants used in this work as well as the associated limit densities $\rho_\mathrm{lim}$ up to which the inequalities of Eq. \eqref{validity} respectively hold are reported in Table \ref{tab_const}. It is worth mentioning that, in the literature, the $p$--wave scattering length varies from 0.45 fm in Refs. \cite{QMCAV8,EFTasym} to 0.84 fm in Ref. \cite{lacroix2}. The adopted value 0.63 fm corresponds to the AV4 interaction (see Ref. \cite{lacroix2}).  
\begin{table}[b]
\caption{Physical values for the $s$--wave scattering length $a_s$, the effective range $r_s$, and the $p$--wave scattering length $a_p$, adopted in the present work. $\rho_\mathrm{lim}$ denotes the density from which each of the conditions of Eqs. \eqref{validity} is violated.}\label{tab_const}
\begin{tabular*}{\linewidth}{@{\extracolsep{\fill}}cccc}
\hline
\hline
                                  &  $a_s$     &   $r_s$   & $a_p$  \\
(fm)                              & --18.9     &   2.75    &  0.63  \\
\hline
$\rho_\mathrm{lim}$ (fm${}^{-3}$) & $5.0\times10 ^{-6}$  & $ 1.6\times10 ^{-3}$    & $ 0.135$  \\
\hline
\hline
\end{tabular*}
\end{table}

The original ELYO functional \cite{ELYO} has been designed by retaining only the $s$--wave terms of Eq. (\ref{LY}), that is by discarding the last term, \textbf{and by requiring that the first relation of Eq. (\ref{validity}) is always satisfied}. The associated validity condition is then generalized to density regimes of interest for nuclear physics by allowing $a_s$ to depend on the density as
\begin{gather} \label{as_rho}
a_s(\rho)=
\left\lbrace
\begin{array}{llc}
 a_s             & \text{if } \rho<\rho_\mathrm{lim}  & \text{(I)} \\
- \Lambda/(3\pi^2\rho)^{1/3}  & \text{if }\rho\geq\rho_\mathrm{lim} &  \text{(II)}
\end{array},\right.
\end{gather}
where $\Lambda \leq 1$ is a control parameter. Throughout this paper, when there is no explicit $\rho$ dependence, $a_s$ refers to the physical value in Table \ref{tab_const}. It is therefore supposed that the $s$--wave scattering length departs from its bare value to approach zero as the density increases, thus modeling in--medium effects.

The ELYO EOS may be mapped with a pure $s$--wave Skyrme mean--field EOS, that is without the $t_2$ gradient component, through a term--by--term identification with respect to the power of $k_F$ (or $\rho$, equivalently) appearing in Eq. \eqref{LY}, leading to
\begin{subequations}\label{condLYs}
\begin{align} 
\label{LYst0}  &t_0(1-x_0)=\dfrac{4\pi\hbar^2}{m}a_s(\rho) ,\\
\label{LYst1}  &t_1(1-x_1)=\dfrac{2\pi\hbar^2}{m}\Bigl[ r_s a_s^2(\rho) +0.19\pi a_s^3(\rho)\Bigr], \\
\label{LYst3}  &t_3(1-x_3)=\dfrac{144\hbar^2}{35m} c_0 (11-2\ln 2)a_s^2(\rho),
\end{align}
\end{subequations}
with $c_0=(3\pi^2)^{1/3}$, provided that the fractional power of the density--dependent $t_3$ term is fixed to $\alpha=1/3$. For pure neutron systems, the above combinations $t_i(1-x_i)$, ${i=0,1,3}$, fully characterize the EOS.

Such a mapped Skyrme--type $t_0-t_1-t_3$ model can be used to describe also SNM. For this, we have not required in Ref. \cite{ELYO} that the very low--density regime of SNM is correctly reproduced. Instead, since the SNM EOS of a Skyrme--type $t_0-t_1-t_3$ model does not depend on the $x_i$'s, the parameters $t_i$ have been fitted so that to impose correct properties close to saturation. The coefficients $x_i$ are then generated by Eqs. (\ref{LYst0})-(\ref{LYst3}) through which they depend on the density.

\begin{figure}[b]
\begin{center}
\includegraphics[width=\columnwidth]{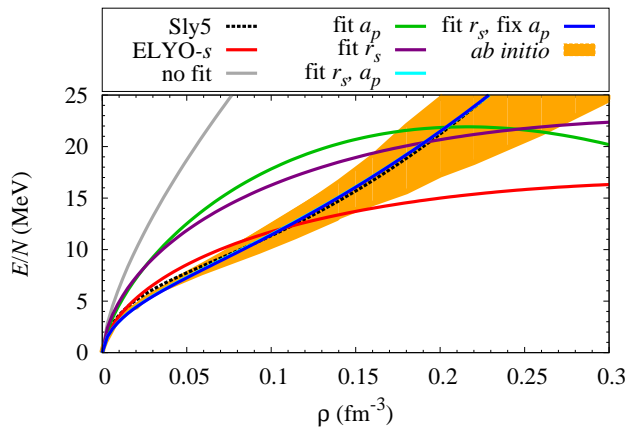}
\caption{Energy per particle of PNM as obtained in the five cases described in the text. Also shown for comparison are the EOS for the ELYO-$s$ (red line) and SLy5 functionals (black dotted line). The orange area contains \textit{ab--initio} results from Refs. \cite{QMCFP,QMCAkmal,NCSMQMC,QMC,QMCGez}. 
The different curves refer to the cases 1-5 discussed in the main text. Note that the case--4 curve (fit $r_s$, $a_p$) (cyan) is practically superposed to the case--5 curve (fit $r_s$, fix $a_p$) (blue).}
\label{fig_EoSPNM}
\end{center}
\end{figure}

The effective range in the region (II) defined by Eq. \eqref{as_rho} is tuned to reach $-4.5$ fm, very different from the bare value (see Table \ref{tab_const}), but required to obtain a reasonable PNM EOS with $\Lambda=1$. The ELYO EOS of PNM, denoted by ELYO-$s$ on Fig.~\ref{fig_EoSPNM} (to emphasize its pure $s$-wave character), depends on this unique phenomenological parameter, whereas the three $t_i$'s are needed to specify the EOS of SNM. The orange area on the figure represents a collection of \textit{ab--initio} results \cite{QMCFP,QMCAkmal,NCSMQMC,QMC,QMCGez} relying on various interactions and many-body methods. Also shown is the PNM EOS produced by the SLy5 Skyrme parametrization \cite{SLY5}, which agrees rather well with the \textit{ab--initio} estimates (we remind the reader that the SLy5 PNM EOS was adjusted on the Akmal et al. EOS \cite{QMCAkmal}) and is, consequently, taken as reference in the present work.

The reason why the readjustment of $r_s$ was a necessary step for the ELYO-$s$ functional may be explained by examining the Lee-Yang formula.  In the case of Ref. \cite{ELYO}, the scattering length is taken equal to $-18.9$ fm up to $k_F \sim$ 0.05 fm$^{-1}$, that is up to the associated $\rho_\mathrm{lim}$. For this value of $k_F$ and using $r_s= 2.75$ fm, $r_s k_F \sim$ 0.14. Up to $k_F \sim$ 0.05 fm$^{-1}$, $|r_sk_F| <1$ and it was checked that the term containing the effective range in the Lee-Yang formula may be safely neglected. Beyond, the scattering length deviates from $-18.9$ fm and it becomes meaningless to keep the associated  value of 2.75 fm for the effective range.  On the other side, if one assumes a prescription such as Eq. (\ref{as_rho}) for the effective range as well and keeps the original ELYO-$s$ functional, then the energy is just proportional to that of a free Fermi gas, as for unitary gases, the proportionality constant being the Bertsch parameter $\xi$. Taking for example $\xi=0.37$, the EOS of PNM is indeed quite well described by such a unitary gas EOS \cite{Ingo} (see also discussion in Ref. \cite{Lac16}). However, an effective range of the form \eqref{as_rho} would lead for ELYO to a value of $\xi$ which is quite different from 0.37, and thus to an EOS for PNM definitely far to be acceptable. If, on the other hand, $r_s$ is kept constant, an additional $k_F^3$ term ($\propto\rho$) survives within the domain (II) and the adjustment of $r_s$ allows for correcting the EOS compared to the case where the bare value of $r_s$ is used (see Fig. 4 of Ref. \cite{ELYO}). The point is that an EOS containing only $s$--wave terms is not sufficient to well describe PNM if both the scattering length and its associated effective range obey a low--density constraint such as Eq. \eqref{as_rho}.

Of course, one may notice that, with $r_s = -4.5$ fm, the relation $|r_sk_F| <1$ is satisfied only at very low densities ($k_F< 0.22$  fm$^{-1}$). This means that, by fitting $r_s$, we renounce the second condition of Eq. \eqref{validity} to be valid in all density regions and maintain only the first (which is, by the way, the only condition strictly required to have a correct EOS at extremely low densities).

Let us now include within the ELYO functional the first $p$--wave contribution which was neglected in the initial version. To this end, we proceed as in Ref. \cite{ELYO}, that is by mapping Eq. \eqref{LY} term--by--term with a Skyrme-like EOS.  The $p$--wave part thus gives rise to a new $t_2$ term related to the $p$--wave scattering length by
\begin{equation} \label{condLYp}
t_2(1+x_2)=\dfrac{4\pi\hbar^2}{m} a_p^3, 
\end{equation}
while Eqs. \eqref{condLYs} still hold. 

Next, we have to choose how to treat the quantities $a_s$, $r_s$, and $a_p$, that is to decide which ones are considered as density dependent through a prescription of the form \eqref{as_rho} and which ones are used as adjustable parameters. Several directions are explored here and one is finally chosen.  Before, let us make two remarks:
\begin{itemize}
\item[$\bullet$] We see from Table \ref{tab_const} that it is actually possible to define four ranges of densities separated by the three values of $\rho_\mathrm{lim}$ associated with each physical constant. Several tests were carried out leading to more or less satisfactory results and it was finally concluded that the best strategy, adopted here,  
consists in considering only two regions: (I) where Eq. \eqref{LY} applies with the conditions of Eqs. \eqref{validity} simultaneously fulfilled; (II) which coincides with the region (II) of Eq. \eqref{as_rho} and starts at the limit density associated with $a_s$ in Table \ref{tab_const}.
\item[$\bullet$] For simplicity, the control parameter $\Lambda$ is assumed to be the same for the three low--density constants, and fixed to unity.
\end{itemize}

As the $s$--wave scattering length dominates, Eq. \eqref{as_rho} is used systematically and we have identified the following cases for which the obtained PNM EOS are represented on Fig. \ref{fig_EoSPNM}:
\begin{itemize}
\item[1 --] no fit (grey curve) -- In addition to Eq. \eqref{as_rho}, we have density dependencies in region (II) for both the effective range
\begin{gather} \label{rs_rho}
r_s(\rho)=
\left\lbrace
\begin{array}{lr}
           r_s              & \text{(I)} \\
1/(3\pi^2\rho)^{1/3}  & \text{(II)}
\end{array}\right.,
\end{gather}
and the $p$--wave scattering length
\begin{gather} \label{ap_rho}
a_p(\rho)=
\left\lbrace
\begin{array}{lr}
           a_p                & \text{(I)} \\
1/(3\pi^2\rho)^{1/3}  & \text{(II)}
\end{array}\right. .
\end{gather}
The minus sign in Eq. \eqref{as_rho}, region (II) disappears for $r_s(\rho)$ and $a_p(\rho)$ as the sign is imposed by continuity argument at the frontier between the two regions.
\item[2 --] fit $a_p$ (green curve) -- As for $r_s$ in the $s$--wave version, the value of $a_p$ in region (II), denoted by $a_p^{\mathrm{II}}$, is adjusted on the SLy5 EOS whereas $r_s(\rho)$ is employed.
\item[3 --] fit $r_s$ (purple curve) -- The treatment of $r_s$ and $a_p$ is interchanged compared to the case 2, that is, we use $a_p(\rho)$ whereas $r_s^{\mathrm{II}}$ is fitted.
\item[4 --] fit $r_s$, $a_p$ (cyan curve)-- Only $a_s$ is density dependent and the values of both $r_s^{\mathrm{II}}$ and  $a_p^{\mathrm{II}}$ are adjusted.
\item[5 --] fit $r_s$, fix $a_p$ to the  physical value in both regions (blue curve) -- Only $a_s$ is density dependent and only the value of $r_s^{\mathrm{II}}$ is adjusted.
\end{itemize}
Case 1 corresponds to a PNM EOS containing only one term, proportional to $\rho^{2/3}$ (see above), and there is no adjusted parameter. In cases 2 and 3 there are extra contributions depending on $\rho^{5/3}$ and  $\rho$, respectively. Each of the latter two cases relies on a single parameter. We observe that these three cases offer a poor reproduction of the SLy5 curve. In contrast, cases 4 and 5 involve three powers of the density. By performing the adjustement of both $r_s$ and $a_p$ for the case 4, we have found a very good agreement with the benchmark EOS with the values $r_s^{\mathrm{II}} =-7.668$ fm and $a_p^{\mathrm{II}}=0.626$ fm.

One may observe that the adjusted value of $a_p$ is indeed very close to its physical value on Table \ref{tab_const} (even if we know that $a_p$ may vary around this value). Owing to this, the optimal choice for us was to retain the option 5, where the physical value of $a_p=0.63$ fm is used and only one parameter is adjusted, $r_s$. The fitted value of $r_s$ in case 5 is $-7.754$ fm, very close to the the value found for case 4.  The two curves corresponding to cases 4 and 5 are practically superposed on Fig. \ref{fig_EoSPNM}. The new version of the functional corresponding to case 5 is denoted as ELYO-$s+p$ in what follows.

\begin{figure}[t]
\begin{center}
\includegraphics[width=\columnwidth]{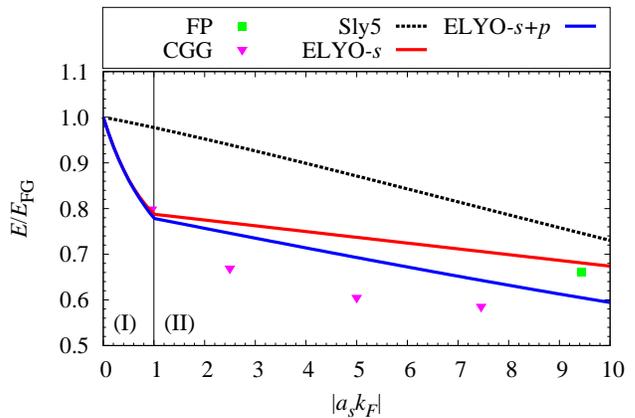}
\vspace*{-0.7cm} 
\caption{Energy of PNM divided by the energy of a free Fermi gas $E_{\mathrm{FG}}$ as a function of $|a_sk_F|$ with $a_s=-18.9$ fm. The colors are the same as in Fig. \ref{fig_EoSPNM}. FP and AV4' stand for the \textit{ab--initio} calculations of Refs. \cite{QMCFP} and \cite{QMCGez,QMCCarl} respectively. $|a_s k_F|=10$ corresponds to a density of $5.10^{-3}$ fm${}^{-3}$ where traditional Skyrme functionals generally break down. (Note that, by mistake, in Fig. 5  of Ref. \cite{ELYO}, the green dot--dashed curve does not correspond to the correct curve and should be replaced by the red curve in the present figure).}
\label{fig_EoSPNM_EFG}
\end{center}
\end{figure}

Expectedly, the $a_p$--terms have very small effects in region (I) (similar discussions were done in Ref. \cite{ELYO}, Sec. III, for $r_s$).  With the adopted value $a_p=0.63$ fm the EOS is continuous at the board $|a_sk_F|=1$, as shown on Fig. \ref{fig_EoSPNM_EFG} which displays the ratio of PNM and free Fermi gas energies as a function of $|a_sk_F|$. To avoid a (very small) discontinuity between regions (I) and (II) we used the value $r_s= -7.754$ fm in both regions (in (I) the contribution related to $r_s$ is anyway negligible). Comparing with \textit{ab--initio} results from Refs. \cite{QMCFP,QMCGez,QMCCarl}, we observe a qualitative improvement when the $p$--wave channel is included.  Note that in all cases the PNM EOS in (I) is given by the Lee-Yang expansion and does not involve any adjustable parameter.

With this adjusted value of $r_s$, it is obvious that the second condition of Eq. \eqref{validity} is valid only at very low densities ($k_F< 0.13$ fm$^{-1}$). We remind that we are not constructing a controlled EFT, but a new type of functional where some parameters are not adjusted, but naturally constrained (such as $a_s$). The optimal description of both SNM and PNM at all densities requires that $r_s$ is treated as a `phenomenological' parameter to adjust.
%
%
%
\section{Finite-size systems and neutron effective mass} \label{sec_drops}
\begin{figure}[b]
\begin{center}
\includegraphics[width=\columnwidth]{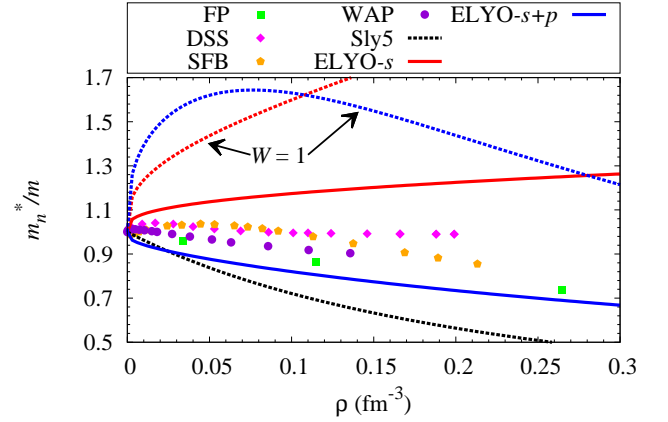}
\caption{Neutron effective masses of the ELYO-$s+p$ functional (blue solid line) compared to the values from SLy5 (black dotted line), the ELYO-$s$ version (red solid line), and \textit{ab--initio} estimates extracted from Ref.~\cite{QMCFP} (FP, green squares), Ref.~\cite{DSS} (DSS, pink diamonds), Ref.~\cite{SFB} (SFB, orange pentagons), and Ref.~\cite{WAP} (WAP, purple circles). The dotted curves for ELYO-$s$ and ELYO-$s+p$ refer to the choice $W_{(1,2)}=1$. }
\label{fig_mstarPNM}
\end{center}
\end{figure}

%
A very important quantity, in particular for the description of finite systems, is the effective mass, defined for PNM as 
\begin{equation}
\dfrac{m^*_n}{m} = \Bigl[ 1 + \dfrac{m}{4\hbar^2} \Theta_n \rho\Bigr]^{-1},
\end{equation}
where $\Theta_n=t_1(1-x_1)+3t_2(1+x_2)$. In virtue of the relations (\ref{condLYs}) and (\ref{condLYp}), $\Theta_n=\frac{2\pi\hbar^2}{m} [r_sa_s^2(\rho) + 0.19\pi a_s^3(\rho) + 6 a_p^3]$ in the case of the ELYO-$s+p$ EDF.  As a result, $m^*_n$ is determined by the effective range, the $s$--wave, and the $p$--wave scattering lengths. In contrast to standard Skyrme EDFs, $\Theta_n$ depends on the density through $a_s(\rho)$. The neutron effective mass computed using the values of $r_s^{\mathrm{II}}$ and $a_p$ (case 5 of the previous section) is plotted on Fig. \ref{fig_mstarPNM} (dotted blue curve) where it is compared to the effective mass associated with the SLy5 and the ELYO-$s$ functionals, as well as to \textit{ab--initio} estimates extracted from Refs. \cite{QMCFP,DSS,SFB,WAP}.

The original ELYO EDF has been recently applied to finite systems composed of neutrons trapped in a harmonic potential \cite{Drops}. This work concluded that the neutron effective mass had to be corrected to get a reasonable reproduction of the droplet energies. The correction was done \textit{via} the introduction of a factor, denoted by $W$, to modulate the $t_1$ contribution and, consequently, $\Theta_n$, while leaving unchanged both EOSs. The effect of this extra parameter is to split the velocity--dependent term Eq. \eqref{LYst1} into the same term weighted by $W$ plus a density--dependent one with the weight $1-W$. The adjustment of $W$ on \textit{ab--initio} neutron drop energies led to the overall reduction of the effective mass observed in Fig. \ref{fig_mstarPNM} from the case $W=1$ (dotted red line) to its optimal value $W=0.396$ (full red line). 

The effective mass ensuing from the inclusion of the $p$--wave term within the ELYO functional (dotted blue curve) is globally as large as in the original ELYO-$s$ case ($W=1$). Accordingly, one can anticipate incorrect droplet energies, which was indeed found. In the same spirit as for the ELYO-$s$ EDF, $m^*_n/m$ must then be corrected. For generality, we therefore introduced two distinct factors $W_{1,2}$ such that Eqs. \eqref{LYst1} and \eqref{condLYp} become:
\begin{subequations} \label{condLYW12}
\begin{align}
& t_1(1-x_1)=W_1\dfrac{2\pi\hbar^2}{m} B_s (\rho), \\
& t_{3'}(1-x_{3'})=(1-W_1) \dfrac{36c_0^2\pi\hbar^2}{10m} B_s (\rho), \\
& t_2(1+x_2)=W_2\dfrac{4\pi\hbar^2}{m} a_p^3, \\
& t_{3''}(1-x_{3''})=(1-W_2)\dfrac{108c_0^2\pi\hbar^2}{5m} a_p^3 ,
\end{align}
\end{subequations} 
where, for compactness, we have set the notation
\begin{equation} \label{Brho}
B_s (\rho) \equiv \Bigl[ r_s a_s^2(\rho) +0.19\pi a_s^3(\rho)\Bigr].
\end{equation}
The extra $t_{3',3''}$ terms have the same density dependencies $\alpha'=\alpha''=2/3$ and, within the EDF, may then be gathered into a unique contribution characterized by $\bar{\alpha}=2/3$, $t_{\bar{3}}= t_{3'} + t_{3''}$ and $x_{\bar{3}}=(t_{3'}x_{3'} + t_{3''}x_{3''})/t_{\bar{3}}$.

Finally, we end up with a generalized Skyrme functional having two $t_3$--like terms, $\{x_i\}_{i=0,1,3,\bar{3}}$ parameters that depend on $\rho$ (by replacing in Eqs. \eqref{condLYs} and \eqref{Brho} $a_s(\rho)$ by $a_s[\rho(\vec{r})]$, $\rho(\vec{r})$ being the local density), and a constant $x_2$. Alternatively, the ELYO-$s+p$ functional may be cast into 
\begin{widetext}
\begin{equation} \label{ELYO2} 
\begin{split}  
\mathcal{E}_c = \mathcal{E}^\mathrm{Sk}_c &- \biggl[X_0 a_s[\rho] + X_3 \rho^{\alpha} a_s^2[\rho] + X_{\bar{3}} \rho^{\bar{\alpha}}D[\rho] \biggr] \biggl[\dfrac{1}{2}\rho^2  - \sum_{q=n,p}\!\!\rho_q^2 \biggr]  \\
&-W_1X_1 B_s[\rho] \biggl[\dfrac{1}{2}\rho\tau+\dfrac{3}{8}(\vec{\nabla}\rho)^2 -\dfrac{1}{4}\vec{J}^{\,2} -\sum_{q=n,p}\!\!\Big(\rho_q\tau_q+\dfrac{3}{4}(\vec{\nabla}\rho_q)^2\Bigr) \biggr]\\ 
& + \quad W_2X_2  a_p^3 \quad\biggl[\dfrac{1}{2}\rho\tau-\dfrac{1}{8}(\vec{\nabla}\rho)^2 - \dfrac{1}{4}\vec{J}^{\,2} + \sum_{q=n,p}\!\!\Big(\rho_q\tau_q-\dfrac{1}{4} (\vec{\nabla}\rho_q)^2\Bigr) \biggr], 
\end{split}
\end{equation}
\end{widetext}
with
\begin{equation}
 D[\rho]=\dfrac{1}{2}(1-W_1)B_s[\rho]+3(1-W_2)a_p^3,
\end{equation}
$B[\rho]$ being the functional extension of Eq. \eqref{Brho}, and
\begin{eqnarray*}
 X_0 &=\dfrac{2\pi\hbar^2}{m}, \quad\quad\; X_1 &=\dfrac{\pi\hbar^2}{2m}, \\
 X_2 &=\dfrac{\pi\hbar^2}{m}, \quad\quad\;  X_3 &=\dfrac{12c_0\hbar^2}{35m}(11-2\ln 2), \\
 X_{\bar{3}}&=\dfrac{3\pi c_0^2\hbar^2}{5m}. \quad &
\end{eqnarray*}
$\tau(\vec{r})$ and $\vec{J}(\vec{r})$ stand for the local kinetic and spin--current densities, respectively (the index $n$ and $p$ refer to the neutron and proton counterparts). The form \eqref{ELYO2} is convenient for neutron drops as the first term $\mathcal{E}^\mathrm{Sk}_c$, defined as the central part of a Skyrme functional with two density-dependent terms (see appendix of Ref. \cite{Drops}) with constant $x_i=1$ (resp. $-1$) for $i=0,1,3,\bar{3}$ (resp. 2), vanishes in that case.

\begin{figure}[!h]
\begin{center}
\includegraphics[width=\columnwidth]{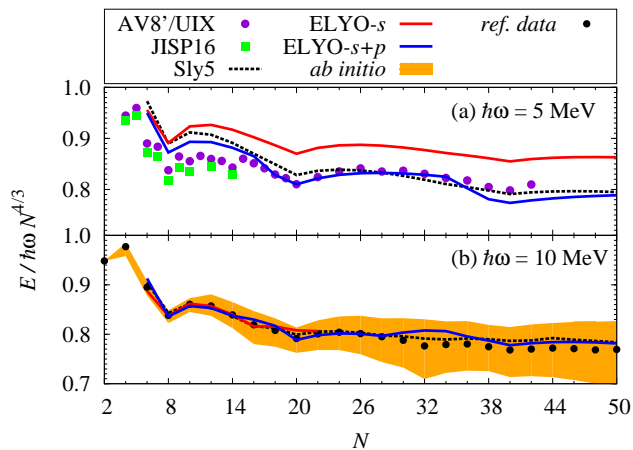}
\vspace*{-0.7cm}
\caption{Energies of neutron drops in traps of frequencies $\hbar\omega = 5$ (a) and 10 (b) MeV, scaled by the Thomas-Fermi approximation $\hbar\omega N^{4/3}$, as obtained with the new ELYO-$s+p$ functional (blue) and compared to values from the ELYO-$s$ (red) and SLy5 (black dotted) EDFs, as well as to \textit{ab--initio} estimates. The purple dots refer to the QMC calculations of Refs. \cite{NCSMQMC,QMC} using the AV8' \cite{AV8} two--body force supplemented by the UIX \cite{UIX} three--body interactions. The green squares indicate results from a configuration--interaction method \cite{NCSMQMC} with the JISP16 force \cite{JISP16}. The orange area on panel (b) represents the collection of these \textit{ab--initio} estimates together with QMC results (using AV8' only or with the IL7 \cite{IL7} three--body interaction), no--core shell--model, and coupled--cluster calculations (with an interaction derived from chiral EFTs \cite{NCSMEFT}). Their average is denoted as ``\textit{ref. data}" (black circles).}
\label{fig_Drops}
\end{center}
\end{figure}

Following Ref. \cite{Drops}, the parameters $W_1$ and $W_2$ are adjusted to reproduce the ``average'' \textit{ab--initio} energies (black dots on Fig. \ref{fig_Drops}) of drops with numbers of neutrons $N=8,12,14,16$, and 20 for a trap frequency $\hbar\omega=10$ MeV. The spin--orbit coupling constant $V_\mathrm{so}$ and the strength $V_\mathrm{pp}$ of a mixed surface/volume pairing interaction are also adjusted. The optimal values of $W_1$, $W_2$, $V_\mathrm{so}$, and $V_\mathrm{pp}$ are reported in Table \ref{tab_res}.
It is worth noticing that the new spin--orbit and pairing coupling constants come naturally closer to the SLy5 values than what was found in the case of ELYO-$s$.

The energies of neutron droplets obtained for $\hbar\omega=5$ and 10 MeV are shown in Fig. \ref{fig_Drops}. We note a considerable improvement for $\hbar\omega = 5$ MeV compared to the pure $s$--wave version of the EDF. Moreover, the numerical instabilities occurring in ELYO-$s$ for $N>22$ in a 10--MeV trap have disappeared.
\begin{figure}[t]
\begin{center}
\includegraphics[width=\columnwidth]{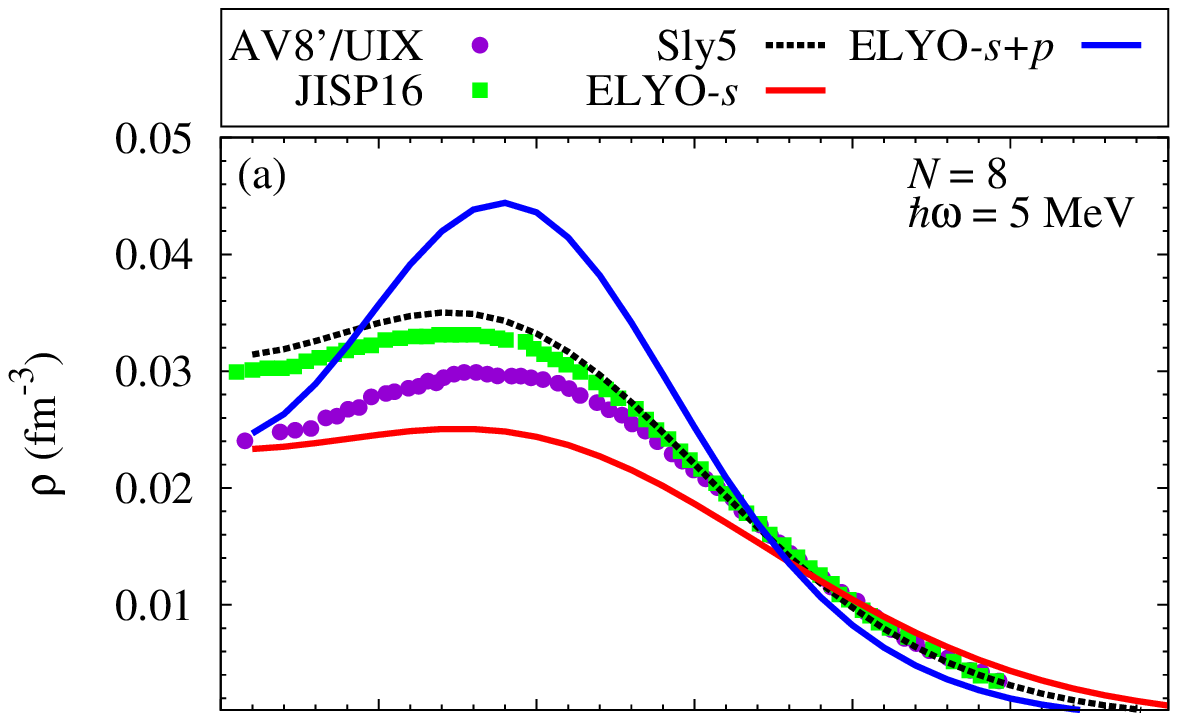}\\
\vspace{-2.13cm}
\includegraphics[width=\columnwidth]{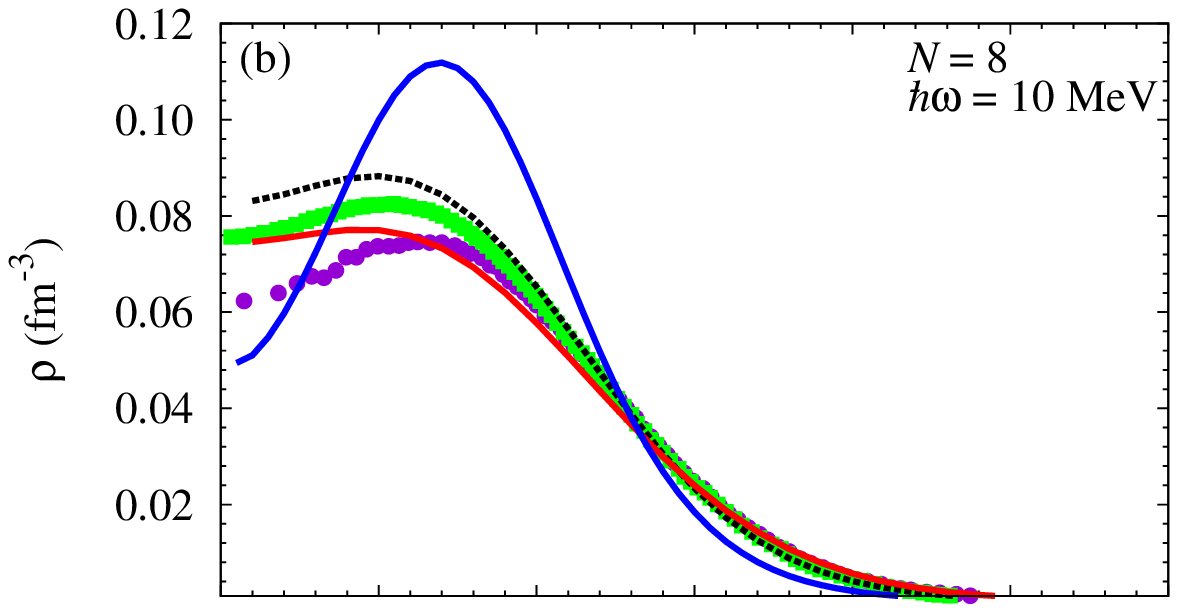}\\ 
\vspace{-2.13cm}
\includegraphics[width=\columnwidth]{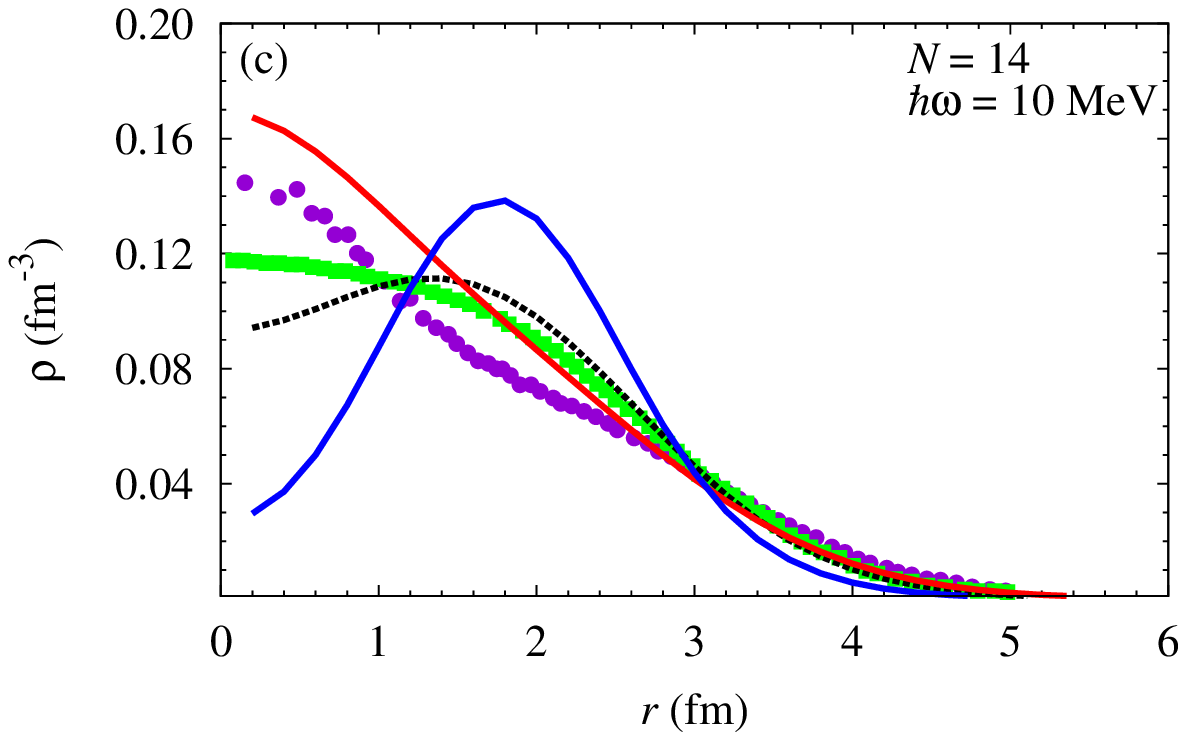}
\vspace*{-0.7cm}
\caption{Density profiles as a function of the distance from the center of the trap obtained with the the Sly5 (black dotted), ELYO-$s$ (red), and ELYO-$s+p$ (blue) functionals for (a) $N=8, \hbar\omega =5$ MeV, (b) $N=8, \hbar\omega =10$ MeV, (c) $N=14, \hbar\omega =10$ MeV.  The \textit{ab--initio} results are extracted from Ref. \cite{NCSMQMC} (purple circles and green squares).}\label{fig_DensDrop}
\end{center}
\end{figure}
\begin{figure}[t]
\begin{center}
\includegraphics[width=\columnwidth]{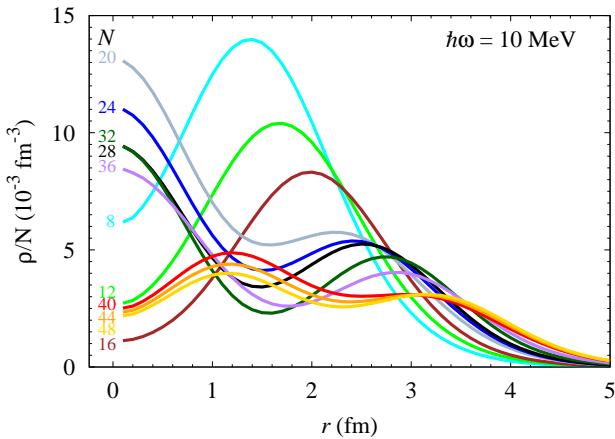}
\vspace*{-0.7cm}
\caption{Densities normalized by the neutron number obtained with the the ELYO-$s+p$ functional for $\hbar\omega =10$ MeV and $8\leq N \leq48$.}
\label{fig_DensDrop_comp}
\end{center}
\end{figure}

To further test the new functional, we plot on Fig. \ref{fig_DensDrop} the density profiles for the systems $N=8$, $\hbar\omega=5$ MeV and $N=8,14$, $\hbar\omega=10$ MeV. While the reference SLy5 curves well agree with the \textit{ab--initio} values, the ELYO-$s+p$ densities exhibit curious bubble--like trends in all the considered cases, entailing a disagreement larger than for the pure $s$--wave EDF.  For illustration, Fig. \ref{fig_DensDrop_comp} compiles the densities normalized by the number of neutrons for drops containing from 8 to 48  neutrons, with $\hbar\omega=10$ MeV.  When $N$ increases from 8 to 16, a strong central depletion (bubble) exists and the peak of density is progressively shifted to the surface of the system. From $N= 20$ to 36 the central depletion disappears and a new one shows up between 1 and 2 fm. Starting from $N=$ 40, a central depletion reoccurs while a second one is located between 2 and 3 fm.

It has been shown that such extremely pronounced bubble structures may be the signature of finite--size instabilities (one may notice in the figure that only this functional predicts these structures) \cite{martini,insta1,insta2}. In Refs. \cite{insta1,insta2}, the linear response theory has been employed to predict this kind of instabilities and then avoid the regions of parameter values responsible for their appearance.  Such an analysis exceeding the scope of this article, the study of the instabilities in EFT--inspired functionals will be presented as the future step of our project.

Note that, for consistency, we could have used reference energies from mean--field calculations with the SLy5 functional, as done for the fit of the other parameters in the previous and following sections, instead of the ``average'' \textit{ab--initio} results defined in Ref. \cite{Drops}.  However, the impact of this choice on the obtained parameters is negligible since both sets of pseudo--data are very close to each other (see Fig. \ref{fig_Drops}).

The neutron effective mass for the optimal values of $W_{1,2}$ is shown in Fig. \ref{fig_mstarPNM}. A strong reduction of $m^*_n$ is observed, yielding values qualitatively comparable to the \textit{ab--initio} estimates of Refs. \cite{QMCFP,DSS,SFB,WAP}.
%
%
%
%
\section{Parameters from symmetric nuclear matter} \label{sec_SNM}
\begin{figure}[t]
\begin{center}
\includegraphics[width=\columnwidth]{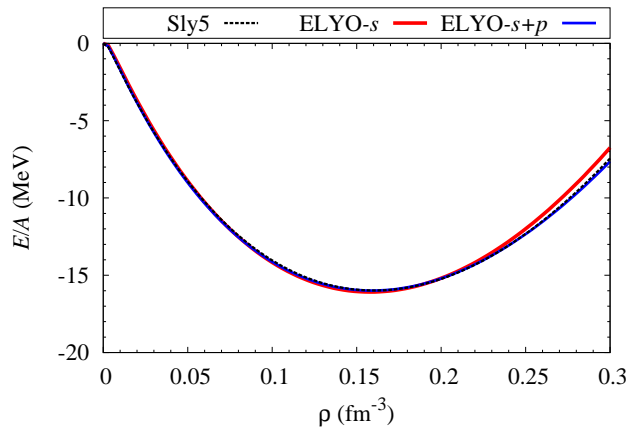} 
\vspace*{-0.7cm}
\caption{ELYO-$s+p$ (blue) SNM EOS compared with those from the original ELYO-$s$ (red) and SLy5 (black dashed line) functionals.}
\label{fig_EoSSNM}
\end{center}
\end{figure}

%
A full definition of the underlying functional requires the determination of all the parameters. The PNM EOS only gives the combinations of Eqs. (\ref{condLYs}) and (\ref{condLYp}). Therefore, as the $\{t_i\}$ and $\{x_i\}$ parameters are still undefined individually, we have to resort to additional constraints. As for for ELYO-$s$ in Ref. \cite{ELYO}, we consider the EOS for SNM that reads:
\begin{equation} \label{EosSNMSkyrme}
 \dfrac{E}{A}=\dfrac{3c_1^2}{5}\dfrac{\hbar^2}{2m}\rho^{2/3} +\dfrac{3}{8} t_0\rho + \dfrac{1}{16}t_3 \rho^{4/3} + \dfrac{3c_1^2}{80} \Theta_s \rho^{5/3},
\end{equation}
where $c_1=(3\pi^2/2)^{1/3}$ and $\Theta_s=3t_1+t_2(5+4x_2)$. Due to Eq. \eqref{condLYp}, $\Theta_s=3t_1+t_2 +\frac{16\pi\hbar^2}{m} a_p^3$ and is constant in contrast to $\Theta_n$ that explicitly depends on the density in region (II). This particular form \eqref{EosSNMSkyrme} in which $W_1$, $W_2$, $t_{3'}$ and $t_{3''}$ are not involved, results from the assumption that the EOS are not affected by the splitting parameters $W_{1,2}$ introduced in the previous section. This is true if and only if we set
\begin{equation}
\begin{split}
& t_{3'} = (1-W_1) \dfrac{9c_1^2}{5} t_1, \\
& t_{3''} = (1-W_2) \dfrac{3c_1^2}{5} t_2(5+4x_2),
\end{split}
\end{equation}
so that these terms respectively recombine with the $t_{1,2}$ components of $\Theta_s$ weighted by $W_{1,2}$. Fitting on the reference SLy5 EOS then provides the values of $t_0$, $t_3$, and $\Theta_s$, the latter of which defining also the isoscalar effective mass
\begin{equation}
\dfrac{m^*_s}{m} = \Bigl[ 1 + \dfrac{m}{8\hbar^2} \Theta_s \rho\Bigr]^{-1}.
\end{equation}
The resulting parameter values are reported in table \ref{tab_res} while   
the obtained EOS is shown on Fig \ref{fig_EoSSNM}. One may notice from Table \ref{tab_res} that such an EOS  corresponds to a high value ($\sim$ 1.4) for $m^*_s/m$ at saturation density that is twice the one of SLy5.

It remains to determine $t_1$, $t_2$ separately, which are important for the detailed form of the functional. Indeed, even though they combine in the EOS of SNM into a unique $\rho^{5/3}$ term and do not impact the results for pure neutron systems, they correspond to different functional form in Eq. \eqref{ELYO2}. For this purpose, one has to resort to extra constraints for the ratio $t_1/t_2$, for instance from the properties of particular nuclei.  This task is out of the scope of the present paper that mainly aims at improving the description of neutron systems (matter and finite--size drops) by including the contribution from $p$--wave scattering.  Accordingly, we report the adjustment of $t_{1,2}$ to a forthcoming work on the application of EFT--guided functionals to nuclei.

The parameters of the ELYO-$s+p$ functional, as well as the associated properties of infinite matter and neutron drops, are summarized on Table \ref{tab_res}. One notices that PNM may be described with only one parameter, as in the ELYO-$s$ case. The number of parameters increases up to 6 (plus the pairing strength) when neutron drops and the neutron effective mass are also considered. By including SNM, the new ELYO functional has finally 9 parameters, that is only one more than the original ELYO-$s$ functional and one less than Skyrme interactions.

\begin{table}[t]
\caption{Summary of the parameters of the ELYO-$s+p$ functional obtained in this work ($\Lambda =1$), organized according to the constraint employed for their fit. The arrows indicate features of the EDF depending on the parameters appearing just above. The index $c$ refers to saturation density. The physical value of $a_p= 0.63$ fm is used.}\label{tab_res}
\begin{tabular*}{\linewidth}{@{\extracolsep{\fill}}lr}
\hline
\hline
\multicolumn{2}{c}{PNM EOS} \\
\hline
\hline
\multicolumn{2}{c}{$r_s^{\mathrm{II}}=-7.754$ fm} \\
\multicolumn{2}{c}{$\Longrightarrow$ Valid Lee-Yang formula} \\
 &  \\
\hline
\hline
\multicolumn{2}{c}{Neutron drops energies and effective mass ($\alpha$=1/3, $\bar{\alpha}=$ 2/3)} \\
\hline
\hline
$W_1 = -0.163$ &  $W_2 = 0.499$ \\
\multicolumn{2}{c}{$\Longrightarrow m^*_n/m|_c = 0.731$} \\
$V_\mathrm{so}=81.20 $ MeV.fm$^5$ & $V_\mathrm{pp}= -252.14$ MeV.fm$^3$ \\
\multicolumn{2}{c}{$\Longrightarrow$ Neutron drop energies} \\
 & \\
\hline
\hline
\multicolumn{2}{c}{SNM EOS} \\
\hline
\hline
$t_0=-1916.910$ MeV.fm$^3$&  $t_3=15344.700$ MeV.fm$^4$ \\
\multicolumn{2}{c}{$\theta_s=-598.97 \text{ MeV.fm}^5$} \\
\multicolumn{2}{l}{$\quad\quad\Longrightarrow\rho_c = 0.159 \text{ fm}^{-3} \quad\quad\quad \Longrightarrow E/A|_c = -15.990$ MeV}  \\
\multicolumn{2}{l}{$\quad\quad\Longrightarrow K_\infty = 223.270 \text{ MeV} \quad\; \Longrightarrow m^*_s/m|_c = 1.403$} \\
\hline
\hline
\end{tabular*}
\end{table}
%
%
%
\section{Application to neutron stars} \label{sec_NS}
%
As a first concrete test of the newly designed ELYO-$s+p$ functional, let us now examine its predictions for the masses and radii of neutron stars. But, before doing this, we discuss the values of the symmetry--energy coefficient calculated at the saturation density, $J$, and its slope $L$, which are both very important for the physics of neutron stars. In the present implemented version of the ELYO functional $J$ and $L$ are respectively equal to 32.96 and 49.13 MeV, very close to the corresponding SLy5 values \cite{dutra}. One may notice that these values are more than reasonable by comparing them with the experimental constraints coming, for instance, from heavy--ion collisions \cite{lattimer}, and from two measurements of the electric dipole polarizability on $^{208}$Pb \cite{lattimer,roca}. Note that the value of $J$ and $L$ we obtained are in the allowed region conjectured from the unitary gas equation of state \cite{Ingo}.

The evolution,  with respect to the Schwarzschild radial distance $r$, of the pressure $P(r)$ and of the mass $m(r)$ enclosed within a sphere of radius $r$ is governed by the Tolman-Oppenheimer-Volkov (TOV) differential equations:
\begin{align}\label{TOV}
& \mathrm{d} m = 4\pi \dfrac{\epsilon}{c^2} r^2 \mathrm{d} r, \\
& \mathrm{d} P = \dfrac{m \epsilon}{c^2r^2}\Bigl(1+\dfrac{P}{\epsilon}\Bigr)\Bigl(1+\dfrac{4\pi P r^3}{mc^2}\Bigr)
  \Bigl(1-\dfrac{2Gm}{rc^2}\Bigr)^{-1}\mathrm{d} r, \nonumber 
\end{align}
where $\epsilon$ stands for the energy density, $G$ for the universal gravitational constant, and $c$ for the speed of light.  Supplemented by the neutron star EOS $(\epsilon(\rho);P(\rho))$, Eqs. \eqref{TOV} form a closed system that can be integrated from a chosen central pressure (or density) up to the boundary condition $P(r=R)=0$, which provides the stellar radius $R$ and the associated mass $M=m(R)$. For the EOS, we adopt here the approximation scheme used in Ref. \cite{NSmodel}. The crust ($\rho<0.076$ fm${}^{-3}$) is described by a compressible liquid--drop model \cite{CDLM} with parameters from the SLy4 Skyrme interaction \cite{SLY5}. The core of the star ($\rho>0.076$ fm${}^{-3}$) is supposed to be composed of a mixture of $n$, $p$, $e^-$, $\mu^-$ at zero temperature. The baryonic contribution to its EOS is given by the asymmetric nuclear matter energy provided by the functional under consideration using the parabolic approximation. The leptonic terms are evaluated in the Fermi gas model for ultra-relativistic $e^-$ and relativistic $\mu^-$. The fractions of each constituent are deduced consistently from the $\beta$-stability conditions.

\begin{figure}[t]
\begin{center}
\includegraphics[width=\columnwidth]{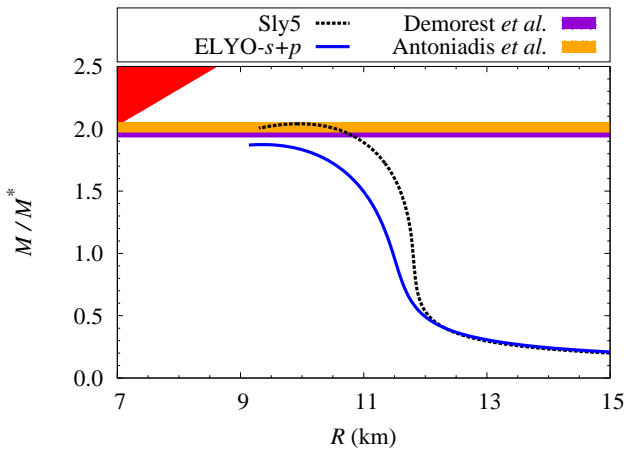} 
\vspace*{-0.7cm}
\caption{Mass-radius diagram of neutron stars determined \textit{via} the TOV equations using the SLy5 (black dashed line) and ELYO-$s+p$ (blue) functionals for the nuclear part of the core EOS. $M^*$ denotes the solar mass. The horizontal bands represent the recent measurements of Demorest \textit{et al.} \cite{Demo} (purple) and Antoniadis \textit{et al.} \cite{Anto} (orange) while the red area is forbidden by general relativity.}
\label{fig_NS}
\end{center}
\end{figure}

Varying the central density between 0.1 and 1.5 fm${}^{-3}$, we obtain the mass-radius relations depicted in Fig. \ref{fig_NS}. The maximal mass $M=2.04$~$M^*$ calculated with the SLy5 EDF coincides with the recent observations of Refs. \cite{Demo,Anto}, represented by the horizontal bands. 
Furthermore, the radius $R=11.6$ km found with SLy5 for an $M=1.4$~$M^*$ neutron star lives well within the expected range $[10.4,12.9]$ km inferred in Ref. \cite{Steiner} (see also the discussion in section II.B of Ref. \cite{Erler}). In contrast, the initial $s$--wave version of the ELYO functional generates negative pressure at high densities, which prevents its use in the TOV equations as the condition for the hydrostatic equilibrium is violated. The same behavior has already been noticed in Ref. \cite{NSmodel} for the D1S Gogny interaction \cite{D1S} that produces a PNM EOS very similar to ELYO-$s$. Accounting for the $p$-wave channel notably improves the description of neutron stars:  The ELYO-$s+p$ EDF yields a maximal mass of 1.88~$M^*$ close to the observed values, and a satisfactory radius of 11.2 km for $M=1.4$~$M^*$.
The value of the maximal mass that we obtain is lower than the one provided by the Skyrme parametrization SLy5 because the EOSs of PNM generated by the two functionals start to be different at densities higher than twice the saturation density. However, it is well known that, at such large densities, it starts to be meaningless to describe the EOS of PNM with a simple EDF--based models and that new degrees of freedom related to the internal structure of nucleons should be explicitly included. This is why we can provide only a kind of qualitative estimation and, for this reason, the fact that our prediction is lower (but not so much anyway) than observations is not for us a crucial issue.
%
%
%
\section{Summary and conclusion} \label{concl}
%
%
This paper proposes an extension of the EFT--inspired ELYO functional. The new ansatz still relies on the Lee-Yang expansion but incorporates the initially neglected first $p$--wave term of the expansion. The PNM may be described with a unique parameter related to the effective range in the Lee-Yang expansion. Considering neutron drops energies and neutron effective, the number of parameters increases to 6, and goes to 9 when SNM saturation properties are included as constraints. Finally, the new ELYO-$s+p$ EDF contains only one parameter more compared to the original $s$--wave version, which is still one less than traditional functionals derived from Skyrme effective forces.

Encouraging results have been obtained:  The description of systems not comprised in the pseudo--data set used for the fit turns out to be significantly improved, as illustrated by the applications done to heavier drops and neutron stars.

Future investigations include first the study of finite--size instabilities, and then the treatment of atomic nuclei with superfluidity effects. This will allow us to further develop and test the functional form and to determine the last parameter needed to fully characterise it, that is the ratio $t_1/t_2$. 
We expect that the analysis of instabilities will slightly change some parameter values. This is why we plan to carry out uncertainty quantification to characterize our calculations only once this study will be completed.

%
%
\section*{Acknowledgment} 
This project has received funding from the European Union's Horizon 2020 research and innovation programme under Grant Agreement No. 654002.

We aknowledge fruitful discussions with Karim Bennaceur and Alessandro Pastore. 

\end{document}